\documentclass{article}

\usepackage[preprint,nonatbib]{neurips_2019}

\usepackage[utf8]{inputenc}
\usepackage[T1]{fontenc}
\usepackage{hyperref}
\usepackage{url}
\usepackage{booktabs}
\usepackage{amsfonts}
\usepackage{nicefrac}
\usepackage{microtype}
\usepackage{graphicx}
\usepackage{amsmath}
\usepackage{float}

\title{Coaching Qwen3 Coder 30B to Think Like a CodeClash Arena Agent}

\author{
  Ivy Ning Zhang \\
  Department of Computer Science \\
  Stanford University \\
  \texttt{para2046@stanford.edu}
}

\begin{document}

\maketitle

\begin{abstract}
Large language model coding agents have recently become useful for software tasks, but weaker or open-weight agents still struggle to reliably interpret user intent and execute complex multi-step workflows. This gap is especially visible in long-horizon settings, where an agent must repeatedly inspect prior outcomes, diagnose failure, and choose the next code edit under interaction constraints. It motivates a natural question: what can we do to improve the thinking process of a weak code agent? We study this question in CodeClash, a code-arena benchmark where the original work evaluates 8 commercial coding agents across 6 arenas through multi-round tournaments. Since Qwen3 Coder Plus ranks last among them, we take the open-weight Qwen3-Coder-30B as a case study and investigate how to improve it with distilled knowledge from stronger agents. Our analysis shows that Qwen3-Coder-30B is not well optimized for arena-style interaction: it frequently produces syntax and protocol-breaking errors and exhibits weak strategic adaptation across rounds. These failures are difficult to correct with vanilla instruction tuning alone, since offline SFT cannot directly verify whether a generated action is valid or beneficial. To address this, we propose ReAct SFT, which rewrites teacher trajectories into explicit \texttt{[obs][thought][act]} chains, and trajectory-quality weighted SFT, which reweights samples to encourage post-edit checking. ReAct SFT substantially improves strategic behavior, and our fine-tuned model outperforms the original Qwen3 Coder Plus in tournament evaluation.
\end{abstract}

\section{Introduction}

In practical coding workflows, we observe a clear gap between weaker and stronger coding agents. Weaker or free tier systems often fail to properly interpret user instructions, lose track of complex requests, or behave unreliably over multi step interactions, whereas stronger frontier agents can handle the same tasks much more consistently. This gap is not fully explained by standard coding benchmarks, which mainly emphasize one shot code generation or issue resolution, such as HumanEval style generation tasks~\cite{chen2021codex} and repository level bug fixing benchmarks like SWE bench~\cite{jimenez2023swebench}. These observations motivate our central question: how can we improve the long horizon behavior of a weak coding agent?

We study this question in \textsc{CodeClash}~\cite{yang2025codeclash}, a benchmark platform for goal oriented, long horizon software engineering. In \textsc{CodeClash}, each model iteratively edits a private repository and then competes with other edited codebases in a code arena, i.e., an executable environment where repository quality is measured through downstream head to head competition rather than direct task completion. The original benchmark evaluates 8 frontier coding agents---Claude Sonnet 4.5, GPT 5, o3, Claude Sonnet 4, GPT 5 Mini, Gemini 2.5 Pro, Grok Code Fast, and Qwen3 Coder---across 6 code arenas through multi round tournaments~\cite{yang2025codeclash}. Among them, Claude Sonnet 4.5 ranks first overall, while Qwen3 Coder ranks last. This motivates our use of the open weight Qwen3 Coder 30B as a posttraining case study, and we use the recorded Claude Sonnet 4.5 trajectories as the teacher source for ReAct SFT supervision. In our experiments, Qwen3 Coder 30B frequently produces syntax and protocol breaking errors and under uses behaviors encouraged by the arena, such as log inspection, outcome analysis, and post edit validation.

A key challenge is that vanilla instruction tuning provides only weak action level control: offline supervised finetuning does not directly tell the model whether a generated action is protocol legal or strategically useful for future rounds. To address this, we build on existing \textsc{CodeClash} trajectories and introduce two forms of supervision. ReAct SFT rewrites strong agent trajectories into explicit \texttt{[Obs][Thought][Act]} annotations while preserving the original bash command, making the competitive decision process more explicit. Trajectory Quality Weighted SFT (TQ-SFT) instead adds a reliability oriented control signal by upweighting trajectories that follow safer modify then check behavior. Empirically, ReAct SFT yields the strongest gains in competitive behavior, while TQ-SFT mainly reduces invalid submissions without producing equally strong tournament improvements.

\section{Related Work}

\subsection{Code arenas and interactive software engineering benchmarks}
Most standard coding benchmarks evaluate models on well specified local tasks, such as code generation from docstrings or issue resolution in a fixed repository context. For example, HumanEval measures functional correctness for one shot code generation~\cite{chen2021codex}, while SWE bench evaluates whether models can resolve real GitHub issues by producing patches that pass repository tests~\cite{jimenez2023swebench}. Interactive coding benchmarks such as InterCode further emphasize execution feedback during problem solving~\cite{yang2023intercode}. In contrast, \textsc{CodeClash} frames coding as goal oriented, long horizon software engineering: agents iteratively edit private repositories and compete in multi round tournaments across executable code arenas~\cite{yang2025codeclash}. This makes \textsc{CodeClash} a natural benchmark for our work, since our goal is not only to improve local code generation, but to improve agent behavior under repeated feedback and competitive pressure.

\subsection{Reasoning and acting supervision for language agents}
Our ReAct style supervision is motivated by prior work on language agents that interleave reasoning and action. ReAct shows that explicitly structuring model outputs as reasoning traces followed by task specific actions can improve sequential decision making and robustness in interactive settings~\cite{yao2023react}. More broadly, recent language agent overviews argue that effective agents require explicit reasoning, grounding in feedback, and action selection under changing context~\cite{su2024languageagents}. These ideas are directly relevant to our setting: in \textsc{CodeClash}, an agent must observe logs and previous outcomes, reason about why a round was won or lost, and then choose the next edit. Our ReAct SFT adapts this perspective to code arenas by rewriting teacher trajectories into explicit \texttt{[Obs][Thought][Act]} supervision.

\subsection{Instruction tuning and its limits in long horizon coding}
Instruction tuning is known to substantially improve model usability and generalization across tasks~\cite{chung2022scaling}. However, open resource studies also show that instruction tuned models acquire different capabilities depending on the supervision data, and that no single instruction tuning recipe uniformly transfers all skills~\cite{wang2023camels}. This is particularly relevant in our setting. Our preliminary experiments suggest that vanilla instruction tuning over arena trajectories can teach Qwen3 Coder 30B to imitate the surface format of agent responses, but not necessarily the deeper observation reasoning action loop required for strong arena play. This motivates our comparison between two more targeted posttraining signals: ReAct style structured supervision, which aims to transfer competitive reasoning, and trajectory quality weighting, which aims to improve reliability under the arena protocol.

\section{Task Setting}

We study weak agent improvement in the \textsc{CodeClash} BattleSnake arena~\cite{yang2025codeclash}. In \textsc{CodeClash}, coding agents iteratively modify a private repository and are evaluated through multi round tournaments. Each round alternates between an edit phase, where the agent interacts with a terminal scaffold under a fixed budget, and a competition phase, where the resulting codebase is executed in the arena. Competition outputs, including logs and round results, are copied back into the repository and become the main source of feedback for the next round. In the BattleSnake setting, the agent must follow a strict single action protocol: at each step it produces exactly one bash command together with a THOUGHT section, receives the execution result, and then continues to the next step.

Our focus is therefore not only whether Qwen3 Coder 30B can generate plausible code, but whether it can behave like an arena agent under repeated feedback. In preliminary experiments, we observe two recurring weaknesses: execution fragility, including syntax and protocol failures, and weak strategic adaptation, including limited use of logs, little post edit validation, and shallow use of previous outcomes. Our goal is to improve both reliability and competitive reasoning in this long horizon setting.

\section{Approach}
\label{sec:approach}
 Our posttraining pipeline begins with standard supervised finetuning baselines, including self play SFT and teacher distilled SFT, and then introduces two more targeted supervision signals: ReAct SFT, which makes the observation reasoning action structure explicit, and Trajectory Quality Weighted SFT (TQ-SFT), which reweights training examples toward safer modify then check behavior.
\subsection{Shared SFT Setup}

Following prior instruction tuning practice~\cite{wang2023camels}, we treat each \textsc{CodeClash} round trajectory as a multi turn conversational finetuning instance. Each trajectory consists of one system message \(s\) and a sequence of user--assistant exchanges. For the \(k\)-th assistant response \(a_k\), we define the history context as the prefix up to the current user turn:
\[
X_k = (s,\; u_1,\; a_1,\; \ldots,\; u_{k-1},\; a_{k-1},\; u_k),
\]
where \(u_k\) denotes the current user-side observation and \(a_k\) is the corresponding assistant output, e.g., a THOUGHT plus one bash action block.

We serialize the full trajectory into a token sequence \(t_{1:L}=(t_1,\ldots,t_L)\). Let \(Y\) denote the set of token positions that belong to assistant spans. We then optimize an assistant only supervised finetuning objective:
\[
\mathcal{L}_{\mathrm{SFT}}
=
-\sum_{j=1}^{L}
\log p_{\theta}(t_j \mid t_{<j})
\cdot
\left\{
\begin{array}{ll}
1, & j \in Y,\\
0, & \text{otherwise.}
\end{array}
\right.
\]
That is, we compute loss only on assistant tokens and mask out system and user tokens. This choice is important because it directly trains the model on the agent outputs we want to improve, rather than on the full conversation transcript.

To enable efficient adaptation of a 30B decoder only model under long context training, we use QLoRA/LoRA style parameter efficient finetuning~\cite{hu2021lora}. In our implementation, the backbone model is loaded in 4 bit quantized form, while forward and backward computation are performed in bfloat16 precision. LoRA adapters with rank \(r=16\) are inserted into the attention projection layers (\texttt{q\_proj}, \texttt{k\_proj}, \texttt{v\_proj}, and \texttt{o\_proj}) of all Transformer blocks, while the original backbone weights remain frozen. We also support both full trajectory training and fixed size windowing over assistant turns, since tournament trajectories can be long while the most relevant dependencies are often local to the recent interaction history.

\begin{figure}[t]
    \centering
    \includegraphics[width=\linewidth]{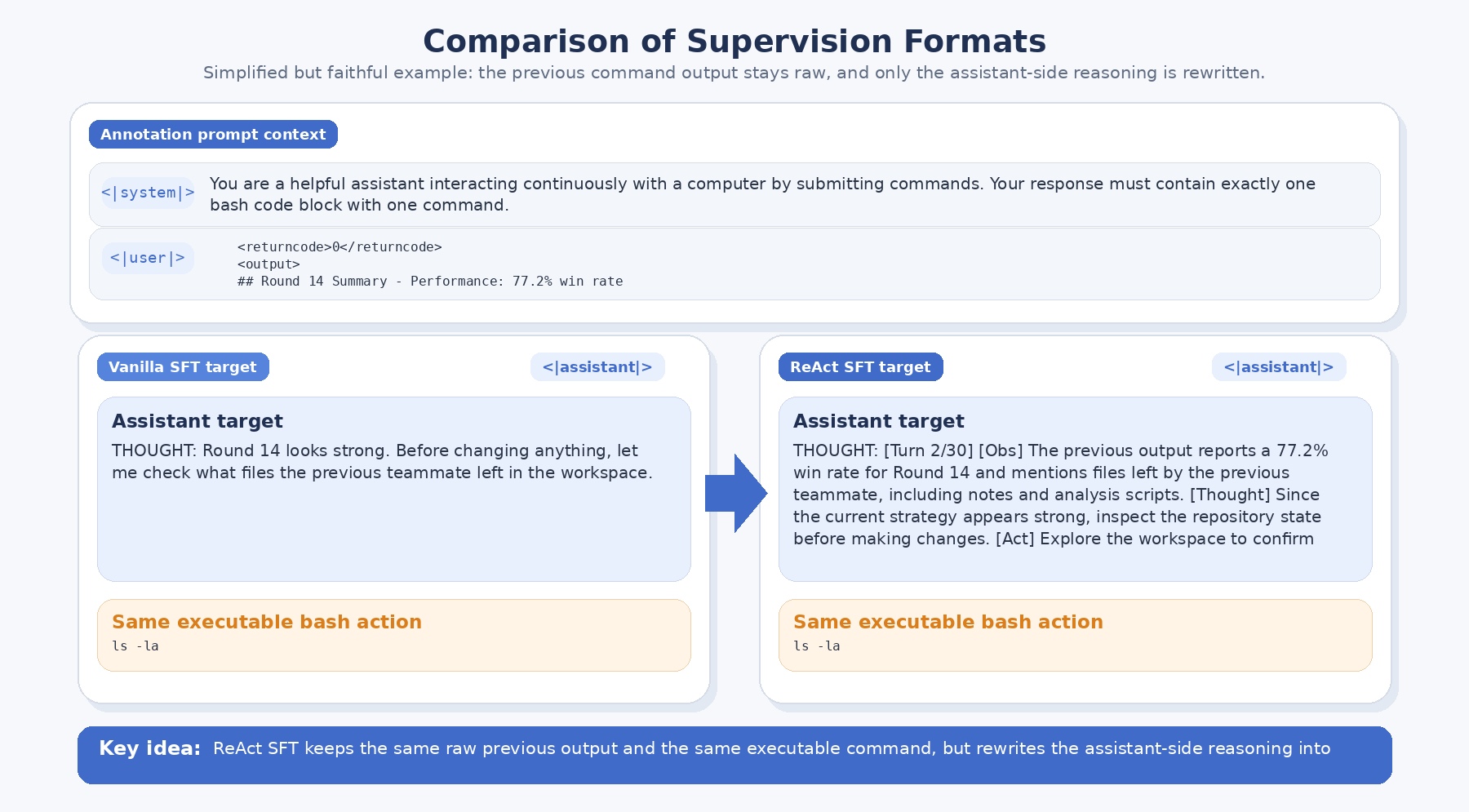}
    \caption{A simplified but faithful example illustrating our supervision rewriting scheme. The raw previous output is preserved in context, but the assistant-side reasoning target is rewritten from an unstructured \texttt{THOUGHT} into explicit \texttt{[Obs][Thought][Act]} fields, while preserving the original executable bash action.}
    \label{fig:react_vs_vanilla}
\end{figure}

\subsection{ReAct SFT}

Our first method addresses a limitation of vanilla instruction tuning: the model can learn to imitate the surface form of an arena response---for example, a plausible THOUGHT followed by a bash command---without learning how to reason strategically under feedback. To make the intended decision process more explicit, we construct ReAct SFT supervision by rewriting teacher trajectories into structured \texttt{[Obs][Thought][Act]} targets. Figure~\ref{fig:react_vs_vanilla} illustrates the difference between vanilla instruction style supervision and our ReAct style reformulation.

Concretely, we use Anthropic's Claude Sonnet 4.5 API to annotate strong agent trajectories step by step; in this setting, the model rewrites its own play trajectories. For each assistant turn, the annotation prompt includes the previous execution output, the current step context, the original THOUGHT text, and the original bash command. The annotator then rewrites only the reasoning portion into three parts: \texttt{[Obs]}, which summarizes the most relevant previous feedback; \texttt{[Thought]}, which states the strategic purpose of the current step; and \texttt{[Act]}, which names the concrete operation being executed. The original bash command is preserved, so the rewritten trajectory remains faithful to the teacher action while making the observation reasoning action structure explicit.

\subsection{Trajectory Quality Weighted SFT (TQ-SFT)}

Our second method targets a different failure mode: invalid or poorly validated edits. In preliminary experiments, Qwen3 Coder 30B frequently makes large modifications without checking whether the code still runs, occasionally removes critical scaffold structure, and submits broken code. This suggests that a useful auxiliary signal is not only what action was taken, but whether the trajectory reflects a more reliable editing process.

To capture this, we construct rule derived trajectory annotations from the top 5 teacher trajectories. Assistant bash commands are mapped into coarse behavior categories such as \texttt{MODIFY\_MAIN}, \texttt{CODE\_CHECK}, \texttt{SUBMIT}, \texttt{STRATEGY\_READ}, and \texttt{STRATEGY\_ANALYSIS}. From the temporal order of these actions, we compute trajectory level statistics such as how often a code modification is followed by checking before submission, and how often the agent submits without validating its changes. These statistics are combined into a heuristic trajectory quality score \(q_i \in [0,1]\), designed to reward modify then check behavior and penalize blind modifications.

During training, this score is converted into a per sample weight \(w_i\), and the supervised loss is reweighted accordingly:
\[
L_{\mathrm{TQ\mbox{-}SFT}}
=
\frac{\sum_i w_i \,\ell_i}{\sum_i w_i},
\]
where \(\ell_i\) is the assistant only cross entropy loss for sample \(i\). Intuitively, trajectories that follow a safer edit validate submit pattern contribute more strongly to optimization, while trajectories that modify code without checking are downweighted. Unlike ReAct SFT, TQ-SFT does not explicitly teach the model how to reason about competition; rather, it biases the model toward more reliable arena behavior.

\section{Experiments}

We evaluate whether posttraining can improve the long horizon arena behavior of Qwen3 Coder 30B in \textsc{CodeClash} BattleSnake. Our experiments compare three forms of posttraining: vanilla self play SFT, ReAct SFT, and Trajectory Quality Weighted SFT (TQ-SFT). Unless otherwise noted, all evaluations are conducted in the same BattleSnake scaffold and follow the original tournament setup of \textsc{CodeClash}~\cite{yang2025codeclash}.

\subsection{Experimental Setup}

All posttraining variants start from the same open weight base model, \texttt{Qwen3 Coder 30B A3B Instruct}, which we refer to as Qwen3 Coder 30B. We compare the following systems:
\begin{itemize}
    \item Vanilla self play SFT: supervised finetuning on default arena trajectories.
    \item ReAct SFT: supervised finetuning on trajectories rewritten into \texttt{[Obs][Thought][Act]} format.
    \item TQ-SFT: supervised finetuning with trajectory quality weighted loss.
\end{itemize}

All finetuned variants use the same shared SFT framework described in Section~\ref{sec:approach}.

\subsection{Training Data}

For Vanilla self play SFT, we use arena trajectories in their original format, where the assistant produces an unstructured \texttt{THOUGHT} followed by a bash command. We consider three data sources in this setting. First, we collect Qwen self play data by running \texttt{Qwen3 Coder 30B} against itself for 10 tournaments, of which roughly 40\% of the resulting round trajectories are valid submissions. Second, we use recorded \texttt{Gemini 2.5 Pro} teacher trajectories, which contribute 1,005 round level trajectories. Third, we use recorded \texttt{Qwen3 Coder Plus} teacher trajectories, which contribute 975 round level trajectories. This setting tests whether standard supervised imitation over arena trajectories, without additional structure, is already sufficient to teach stronger arena behavior.

For ReAct SFT, we construct training data from stronger teacher trajectories by rewriting each assistant step into structured \texttt{[Obs][Thought][Act]} supervision while preserving the original bash command. Our main teacher source is the recorded Claude Sonnet 4.5 BattleSnake corpus. We use Claude Sonnet 4.5 itself as the annotator for step level rewriting. In the initial ablations over context length, we use an early rewritten subset of about 150 trajectories. After identifying the most promising setting, we expand the rewritten corpus to up to 430 trajectories and use the larger set to train the final 6 turn model.

For TQ-SFT, we build a teacher dataset from the strongest recorded agents, including Claude Sonnet 4.5, GPT 5, GPT 5 Mini, Claude Sonnet 4, and Gemini 2.5 Pro. Together, these five agents contribute 4,905 completed round level BattleSnake trajectories. We annotate each trajectory with rule derived action labels and trajectory quality scores computed from edit/check/submit patterns, and convert these scores into per sample loss weights during training. The purpose of this variant is to encourage safer modify then check behavior rather than directly teaching stronger competitive strategy.

\subsection{Evaluation}

We use the \textsc{CodeClash} framework to benchmark model performance in the BattleSnake arena. Performance is measured at both the round and tournament levels. At the round level, we report non tie win rate across 1,000 simulator runs with different random initializations. At the tournament level, we compare cumulative outcomes across all 15 rounds, including total score, round wins, and tie rounds. In addition to competitive metrics, we also track reliability oriented outcomes such as invalid submission rounds and round collapse behavior.

\subsection{Experimental Details}

All experiments use QLoRA as described in Section~\ref{sec:approach}. We set the learning rate to \(1\times10^{-5}\), use batch size \(1\) with gradient accumulation of \(8\), and compute loss only on assistant tokens with sample level normalization so that longer trajectories do not dominate training. Because \textsc{CodeClash} trajectories are long and GPU memory is limited, we train on sliding windows of \(K\) assistant turns together with their corresponding user observations.

For ReAct SFT, we experiment with context windows of \(K \in \{2,4,6,8\}\) in order to identify the most effective context length. For TQ-SFT, we use \(K=6\). We keep the optimization setup fixed across variants so that differences are driven primarily by supervision structure and data quality rather than by training hyperparameters.

\subsection{Preliminary Baselines}
Before introducing ReAct SFT and TQ-SFT, we ran a set of preliminary SFT experiments using self play data and several teacher sources. Table~\ref{tab:preliminary_baselines} summarizes these comparisons and reveals a consistent instability in ordinary arena trajectory finetuning. Although vanilla SFT can yield a modest win rate improvement when trained on Qwen's own self play data, it does not reliably teach the deeper interaction loop required for competitive adaptation. Instead, the resulting models remain vulnerable to the structural failures discussed above, including invalid submissions caused by syntax corruption and removal of the BattleSnake starter block. This instability becomes more severe when the student is trained directly on other agents' trajectories: teacher data SFT using \texttt{Qwen3 Coder Plus} or \texttt{Gemini 2.5 Pro} frequently collapses into all tie rounds or invalid execution states, suggesting that naive imitation can amplify behavior drift rather than transfer robust strategy. See Appendix Section~\ref{sec:appendix_failure_analysis} for a detailed error analysis.

We observe a similar pattern in a self play variant of TQ-SFT; this result is discussed in Section~\ref{sec:results}. Together, these results suggest that vanilla style SFT may help the model adapt to the surface form of its own trajectories, but does not reliably improve strategy and can even increase failure risk at inference time.
\begin{table*}[t]
\centering
\small
\begin{tabular}{lllcccc}
\toprule
Player A & Player B & Winner & Win & Lose & Tie & Primary failure side \\
\midrule
Qwen3-30B SFT (self-play) & Qwen3-Coder-30B base & Player A & 8 & 2 & 0 & both \\
Qwen3-30B SFT (qwen-coder-plus) & Qwen3-Coder-30B base & Player B & 1 & 4 & 0 & mostly A \\
Qwen3-30B SFT (gemini-2.5-pro) & Qwen3-Coder-30B base & Player A & 2 & 1 & 0 & both \\
\bottomrule
\end{tabular}
\caption{Preliminary baseline comparisons at the round level, excluding abnormal rounds with outcomes $(0,0,1000)$, $(0,1000,0)$, or $(1000,0,0)$. Winner is determined by the higher number of round wins among valid rounds. Primary failure side indicates whether failed rounds were mainly caused by Player A, Player B, or both sides producing invalid submissions or structural runtime errors.}
\label{tab:preliminary_baselines}
\end{table*}

\section{Results and Analysis}

\subsection{Main Tournament Results}
\label{sec:results}

We first explored ReAct style supervision with context windows of 2, 4, and 8 turns using an initial rewritten subset of about 150 trajectories. Table~\ref{tab:ablation_battlesnake_results} shows that the 4 turn variant performed best among these early ablations. ReAct 2 turn appears too short to preserve enough arena history for strategic adaptation, while ReAct 8 turn appears too long and noisy, making it harder for the student to extract stable decision patterns.

By contrast, TQ-SFT is better understood as a reliability oriented baseline than a strong strategic one. It improves some robustness related behaviors, but it still underperforms ReAct 4 turn in direct competition. In a self play variant, TQ-SFT reduced invalid submissions in the generated data from roughly 40\% to 25\%, yet the resulting checkpoint failed more often in actual tournament play. This suggests that imitation based finetuning alone does not reliably improve strategic behavior and may even increase deployment time failures through behavioral drift.

Motivated by the strong performance of ReAct 4 turn, we then expanded the rewritten corpus to up to 430 trajectories and trained a 6 turn ReAct variant as our final model. Table~\ref{tab:main_battlesnake_results} reports the resulting main tournament comparisons. ReAct 6 turn wins decisively against both \texttt{Qwen3 Coder 30B base} and \texttt{Qwen3 Coder Plus}, while still remaining clearly below Claude 4.5. Compared with the earlier ReAct variants and TQ-SFT, the final model also shows fewer abnormal collapse rounds, further supporting the value of structured observation reasoning action supervision for competitive code editing agents.

\begin{table*}[t]
\centering
\small
\begin{tabular}{lllcccc}
\toprule
Player A & Player B & Winner & Win & Lose & Tie & Primary failure side \\
\midrule
ReAct 2 turn & Qwen3 Coder 30B base & Player B & 1 & 12 & 0 & both \\
ReAct 8 turn & Qwen3 Coder 30B base & Player B & 1 & 12 & 1 & both \\
ReAct 4 turn & Qwen3 Coder 30B base & Player A & 6 & 2 & 1 & both \\
TQ-SFT & Qwen3 Coder 30B base & Player B & 3 & 6 & 3 & mostly B \\
ReAct 6 turn & TQ-SFT & Player A & 11 & 1 & 2 & both \\
TQ-SFT & Vanilla SFT (self play) & Player A & 3 & 1 & 0 & mostly B \\
ReAct 4 turn & TQ-SFT & Player A & 8 & 3 & 3 & both \\
\bottomrule
\end{tabular}
\caption{Ablation results across post-training variants in BattleSnake, excluding abnormal rounds with outcomes $(0,0,1000)$, $(0,1000,0)$, or $(1000,0,0)$ from the Win/Lose/Tie counts. Winner is determined by the higher number of round wins among valid rounds. `Primary failure side` summarizes which side mainly caused the excluded abnormal rounds.}
\label{tab:ablation_battlesnake_results}
\end{table*}

\begin{table}[t]
\centering
\small
\begin{tabular}{lllcccc}
\toprule
Player A & Player B & Winner & Win & Lose & Tie & Primary failure side \\
\midrule
ReAct 6 turn & Qwen3 Coder 30B base & Player A & 9 & 3 & 1 & both \\
ReAct 6 turn & Qwen3 Coder Plus & Player A & 12 & 0 & 0 & mostly B \\
ReAct 6 turn & Claude 4.5 & Player B & 0 & 15 & 0 & none \\
\bottomrule
\end{tabular}
\caption{Main tournament comparisons for the strongest model, ReAct 6-turn, excluding abnormal rounds with outcomes $(0,0,1000)$, $(0,1000,0)$, or $(1000,0,0)$ from the Win/Lose/Tie counts. Winner is determined by the higher number of round wins among valid rounds.}
\label{tab:main_battlesnake_results}
\end{table}

\subsection{Discussion}

ReAct SFT improves competitive performance primarily by changing the interaction process rather than only the output format. The ReAct trained model is more likely to revisit round results, inspect logs, and choose edits tied to observed failure modes. This is consistent with the gains of the 4 turn and 6 turn variants and suggests that structured supervision transfers part of the observation reasoning action loop used by stronger agents.

TQ-SFT improves a different dimension. By upweighting trajectories with safer modify then check behavior, it reduces invalid and protocol breaking actions and raises the floor of performance. However, it does not directly teach the model which edits are strategically useful, so the model often becomes more cautious without becoming much more adaptive. Table~\ref{tab:behavior_compare} provides a partial behavioral breakdown consistent with this interpretation: ReAct SFT shifts behavior toward round aware diagnosis and testing, whereas TQ-SFT mainly increases checking and analysis behaviors associated with reliability. Detailed behavioral analysis of the tournament runs is provided in Appendix Section~\ref{sec:appendix_failure_analysis}.

\begin{table*}[t]
    \centering
    \small
\begin{tabular}{lccccc}
    \toprule
    Action category & Base Qwen & Qwen Coder Plus & Claude 4.5 & ReAct SFT & TQ-SFT \\
    \midrule
    Round / result checks (A) & 1.22\% & 4.99\% & 31.75\% & 21.51\% & 9.93\% \\
    Edits-log checks (B) & 0.00\% & 0.07\% & 1.67\% & 6.69\% & 0.00\% \\
    Simulation slice checks (C) & 0.00\% & 0.40\% & 18.28\% & 0.00\% & 0.00\% \\
    Code structure checks (D) & 6.94\% & 14.88\% & 8.08\% & 1.16\% & 0.97\% \\
    Local testing (E) & 4.90\% & 0.35\% & 7.01\% & 11.34\% & 8.72\% \\
    Analysis scripts (F) & 0.00\% & 0.29\% & 11.98\% & 9.88\% & 12.59\% \\
    \bottomrule
\end{tabular}
    \caption{Partial behavioral action breakdown across representative agents in BattleSnake. Percentages denote the share of assistant bash actions assigned to the selected categories shown here, so columns do not sum to 100\%. ReAct SFT shifts behavior toward round-aware diagnosis and testing, moving closer to Claude 4.5, while TQ-SFT mainly increases testing and scripted analysis associated with reliability.}
    \label{tab:behavior_compare}
\end{table*}

\section{Conclusion and Future Work}

We studied whether posttraining can improve a weaker open weight coding agent in a long horizon competitive coding benchmark. Using \texttt{Qwen3 Coder 30B} in \textsc{CodeClash} BattleSnake, we found that the base model is limited by both execution fragility and weak arena specific reasoning over logs, prior outcomes, and iterative feedback. Across the posttraining variants we evaluated, ReAct style supervision was the most effective: the final 6 turn model, trained after expanding the rewritten corpus beyond the initial ablation subset, substantially outperformed the base model and clearly exceeded TQ-SFT in tournament play.

Overall, our results suggest that improving coding agents for sequential software tasks requires more than standard instruction tuning or reliability oriented filtering. What matters is transferring a feedback driven action policy that helps the model connect observations, reasoning, and edits over repeated rounds. Structured posttraining is therefore a promising direction for open weight coding agents, although substantial robustness gaps still remain. As future work, it would be valuable to compare other open weight models of similar scale and to study stronger posttraining signals that combine strategic reasoning with explicit validation.

\section*{Acknowledgements}
We thank Chenglei Si for guidance throughout this work and for helping us obtain the original tournament trajectories collected by CodeClash author John Yang, which made these experiments possible.

\bibliographystyle{unsrt}
\bibliography{references}

\appendix
\onecolumn
\section{Appendix}
\label{sec:appendix}
\subsection{Failure Analysis of Vanilla and Self-Play SFT Baselines}
\label{sec:appendix_failure_analysis}

As shown in Table~\ref{tab:preliminary_baselines}, the self-play SFT model improves over the base model in several rounds, but both models still produce multiple 0v0 all-tie rounds caused by invalid submissions or service failures. Table~\ref{tab:qwen_failures} shows that the dominant failure modes include accidental removal of the startup block (e.g., \texttt{run\_server}) and structural corruption (e.g., \texttt{return outside function} and broken \texttt{def move} signatures). A system-level safeguard reduces startup-block removal but does not eliminate instability, since aggressive structural edits such as large-range \texttt{sed} commands or full file overwrites can still introduce invalid code. Distillation from stronger frontier models (Claude-4/4.5) and finetuning on other agent trajectories (\texttt{qwen-coder-plus}, \texttt{gemini-2.5-pro}) show similar structural failure patterns.

\begin{figure}[H]
    \centering
    \includegraphics[width=\columnwidth]{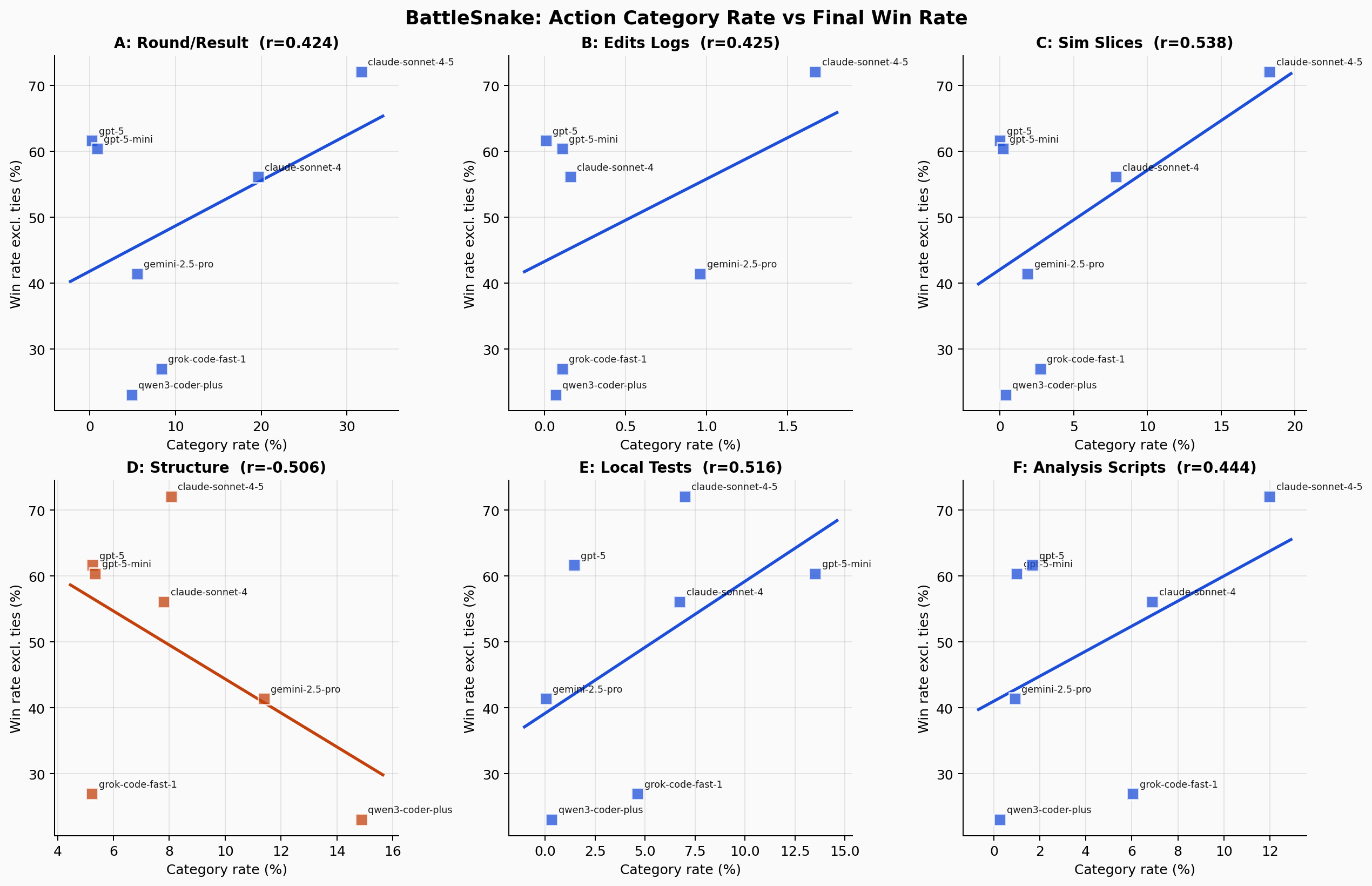}
    \caption{Correlation between error-checking and diagnosis behavior and win rate in the BattleSnake arena.}
    \label{fig:action_vs_winrate_battlesnake}
\end{figure}

Behaviorally, Qwen-based agents perform fewer structured diagnostic actions (Table~\ref{tab:action_categories} and Figure~\ref{fig:action_vs_winrate_battlesnake}) and tend to apply edits without sufficient intermediate validation. Large-range modifications, such as full overwrites or truncate-and-append patterns, frequently disrupt code structure and increase the risk of syntax and runtime errors. This combination of limited diagnostic effort and high-risk editing strategies reduces robustness in tournament settings, particularly for Qwen3-30B.

These results suggest that improving tournament performance requires not only stronger supervision but also explicit behavior-level safeguards, such as pre-submit \texttt{py\_compile} checks and startup-block validation, together with more systematic diagnostic procedures. Instruction tuning alone can shape response patterns, but it does not ensure runtime validity. This motivates our shift from response-level imitation toward learning action-level validity, so that edits remain executable and structurally safe within the deployment scaffold.

\begin{table}[t]
\centering
\small
\caption{Most frequent failure modes in Qwen self-play and the corresponding agent actions that triggered them, ordered by empirical frequency.}
\label{tab:qwen_failures}
\begin{tabular}{p{3.2cm} p{5.2cm}}
\toprule
\textbf{Observed Failure} & \textbf{Typical Agent Action Trigger} \\
\midrule
Startup entrypoint removed \newline
(\texttt{run\_server} block missing)
& Full-file overwrite of \texttt{main.py}
(\texttt{cat > main.py}) or truncate-to-EOF edits
(\texttt{sed -i 'x,\$d' main.py}) \\

Syntax / structural corruption \newline
(e.g., \texttt{return outside function})
& Large-range deletion plus append replacement; partial function rewrite with indentation drift \\

Invalid metadata response \newline
(\texttt{invalid character '<'})
& File rewrite that compiles but causes the server to return malformed or non-JSON output \\
\bottomrule
\end{tabular}
\end{table}

\begin{table}[t]
\centering
\small
\caption{Definitions, purposes, and training value of agent-side diagnostic action categories.}
\label{tab:action_categories}
\begin{tabular}{
p{0.8cm}
p{3.2cm}
p{3.2cm}
p{3.6cm}
}
\toprule
\textbf{Action} & \textbf{Representative Actions} & \textbf{Primary Purpose} & \textbf{Training / Evaluation Value} \\
\midrule
A & \texttt{ls -la /logs/rounds/} \newline
\texttt{cat results.json}
& Confirm win/loss outcomes and anomalous rounds
& Builds the outcome-feedback loop and avoids blind edits \\

B & \texttt{ls -la /logs/edits/} \newline
\texttt{tail -n ...}
& Inspect edit trajectories and submit status
& Connects what changed with what happened \\

C & \texttt{cat sim\_<id>.jsonl | tail -n 5}
& Inspect failed or critical simulations
& Localizes crash and timeout patterns for targeted fixes \\

D & \texttt{grep -n "def move" main.py}
& Verify critical code structure
& Prevents entrypoint or signature corruption \\

E & \texttt{pytest} \newline
\texttt{python test\_*.py}
& Validate logic before submission
& Shifts failures left to local checks \\

F & \texttt{python analyze\_round.py}
& Aggregate multi-game patterns
& Enables reusable cross-round analysis \\
\bottomrule
\end{tabular}
\end{table}

\end{document}